\documentclass[aps,prb,twocolumn,showpacs,amsmath]{revtex4}
\usepackage{graphicx}
\usepackage{bm}

\begin{document}

\title{Field and current distributions and ac losses in a bifilar stack of superconducting strips}

\author{John R.\ Clem}
\affiliation{%
   Ames Laboratory and Department of Physics and Astronomy, \\
   Iowa State University, Ames, Iowa, 50011--3160}

\date{\today}

\begin{abstract} 
In this paper I first analytically calculate the magnetic-field and sheet-current distributions generated in an infinite stack of thin superconducting strips of thickness $d$, width $2a \gg d,$ and arbitrary separation $D$ when adjacent strips carry net current of magnitude $I$ in opposite directions.  Each strip is assumed to have uniform critical current density $J_c$, critical sheet-current density $K_c = J_c d$, and critical current $I_c = 2a K_c$, and the distribution of the current density within each strip is assumed to obey critical-state theory.  I then derive  expressions for the ac losses due to magnetic-flux penetration both from the strip edges and from the top and bottom of each strip, and I express the results in terms of integrals involving the perpendicular and parallel components of the magnetic field.  After numerically evaluating the ac losses for typical dimensions, I present analytic expressions from which the losses can be estimated.
\end{abstract}

%PACS 2006
%74.25.Ha 	Magnetic properties
%74.25.Nf 	Response to electromagnetic fields (nuclear magnetic resonance,
%surface impedance, etc.)
%74.25.Op 	Mixed states, critical fields, and surface sheaths
%74.25.Sv 	Critical currents
%74.78.Bz 	High-Tc films
\maketitle

\section{Introduction} %***** 
The magnetic-field and sheet-current distributions generated in an infinite stack of superconducting strips, all carrying current in the same direction, were calculated analytically in Refs.\ \onlinecite{Mawatari97a} and \onlinecite{Muller97} using an extension of a method first introduced by Mawatari.\cite{Mawatari96}  The results were then put to use to calculate the hysteretic ac losses.  Such calculations can be applied to estimate the ac losses in pancake coils wound from long lengths of second-generation (2G) high-temperature superconducting tapes.\cite{Polak06,Grilli07}  

Recently noninductive coils with bifilar windings (in which adjacent tapes carry current in {\it opposite} directions) have been fabricated using 2G superconducting tapes for use in superconducting fault-current limiters (SCFCLs).\cite{Verhaege99,Ahn06,Xie07,Kudymow07,Kraemer08}  
References \onlinecite{Noe07} and \onlinecite{Malozemoff08} give excellent reviews of this topic.  Each tape consists of a superconducting film  of the order of 1 $\mu$m in thickness and 1 cm in width.  A thin insulating buffer layer separates the film  from the underlying metallic base, which is typically tens of $\mu$m in thickness.  Usually surrounding this structure is a normal-metal (e.g., copper) stabilizer, such that the total tape thickness is a fraction of 1 mm.   Accounting for the  thickness of the insulation between tapes, the spacing $D$ between the superconducting films in adjacent tapes is of the order of a few mm. Since the interleaving tapes in such coils carry current in opposite directions,  the current-generated magnetic fields are localized within the windings and decay very rapidly outside the coil.

To determine the current distribution within a bifilar stack of superconducting strips is not  trivial.  Roughly speaking, when a strip carries an increasing current, the current density just slightly exceeds the the critical current density $J_c$ only at the strip edges, where vortices penetrate  and carry perpendicular magnetic flux through the strip.  The middle portions carry a current density less than $J_c$, and therefore no perpendicular magnetic flux penetrates the strip there.  However, there is a magnetic field parallel to the strip, and vortices can enter the top and bottom of the strip, remaining nearly parallel to the surfaces. 

In Sec. II, I present solutions for the current-density and magnetic-field distributions in an infinite stack of superconducting strips, each strip carrying a current of magnitude $I$ but adjacent strips carrying currents in opposite directions. In Sec. III, I show how to calculate the hysteretic ac losses generated by magnetic flux penetrating both from the strip edges and from the top and bottom surfaces.  I give a brief summary and discuss the results in Sec. IV.

\section{Infinite bifilar stack}

As in many earlier calculations of the properties of thin-film
superconductors, I consider only high-$\kappa$ type-II superconductors and 
assume for simplicity that the magnitude of the self-field ${\bm H}$ at the film
edges or top and bottom surfaces is typically much larger than
$H_{c1}$, such that the magnetic induction in the superconducting film is
given to good approximation by ${\bm B} =
\mu_0 {\bm H}$.  For small currents this assumption leads to an overestimate of both the degree of magnetic flux penetration and the corresponding ac losses. I also will treat the quasi-static penetration of vortices into the film using critical-state theory,\cite{Campbell72} parameterized by a critical current density $J_c$ that is independent of the local magnetic induction.

\begin{figure}%***** Fig.1 ************************
\includegraphics[width=8cm]{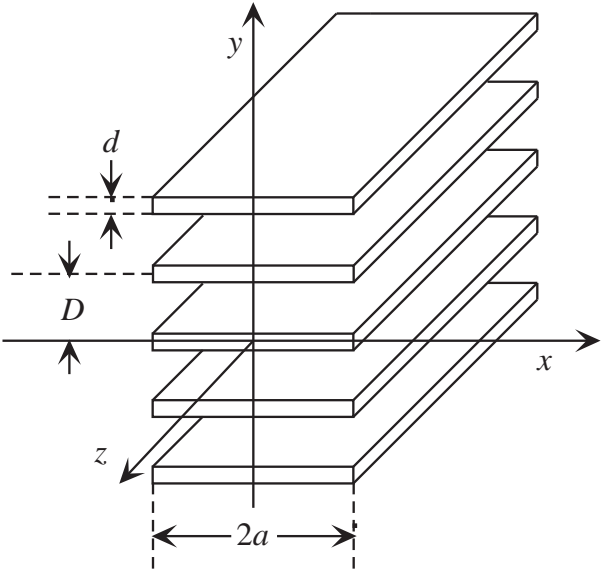}
\caption{Infinite bifilar stack: a stack of thin ($d \ll 2a$) superconducting strips of infinite length
in the
$z$ direction, with those at $y = 0, \pm 2D, \pm 4D, ...$  carrying  current  $I$ in the $+z$ direction and those at $y = \pm D, \pm 3D, ...$ carrying the same current in the $-z$ direction.  }
\label{Fig01}
\end{figure} 

Figure 1 shows the film geometry under consideration, an infinite stack of superconducting strips of width $2a$, thickness $d \ll 2a$, and infinite length parallel to the $z$ axis, equally spaced along the $y$ axis with separation $D$. The strips are assumed to be identical, characterized by a uniform critical current density $J_c$, critical sheet-current density $K_c = dJ_c$, and critical current $I_c = 2 ad J_c = 2a K_c$.  

Since the magnetic-field and current-density distributions depend only upon the coordinates $x$ and $y$ it is convenient to describe the magnetic field outside the strips as ${\cal H}(\zeta)=H_y(x,y)+iH_x(x,y)$, which is an analytic function of the complex variable $\zeta=x+iy$.
The Biot-Savart law for the complex 
field~\cite{Clem73,Mawatari01,Brojeny02} 
can be expressed as 
\begin{equation}
   {\cal H}(\zeta)= \frac{1}{2\pi} 
     \int_{-a}^{+a}du \sum_\infty^\infty\frac{K_{zn}(u)}{\zeta-u-inD} , 
\label{Biot}
\end{equation} where $K_{zn}(x)$ is the sheet-current density in layer $n$ centered at $(x,y) = (0,nD)$.  The currents in a bifilar stack are distributed such that for $n = 0,\pm 2, \pm 4, \pm 6,...,$ the sheet-current density is $K_{zn}(u) = K_z(u)$, the same as in the layer $n=0$, and for $n = \pm 1, \pm 3, \pm 5,...,$ the sheet-current density is $K_{zn}(u) = -K_z(u)$.  The resulting sum in Eq.\ (\ref{Biot}) can be evaluated, yielding
\begin{eqnarray}
{\cal H}(\zeta)\!\!&=&\!\! \frac{1}{2D} 
     \int_{-a}^{+a}\!du \frac{K_z(u)}{\sinh[\pi(\zeta-u)/D]}\\
&=& \!\!\frac{1}{2D} 
     \int_{-a}^{+a}\!\!\!du \frac{K_z(u)\sinh(\pi\zeta/D)\cosh(\pi u/D)}{\sinh^2(\pi\zeta/D)-\sinh^2(\pi u/D)},
\end{eqnarray}
where the second expression follows from the symmetry that $K_z(-u)=K_z(u)$.
Note that the complex field has the desired properties that ${\cal H}(\zeta+iD)=-{\cal H}(\zeta)$ [i.e., $H_x(x,y + D) = -H_x(x,y)$ and $H_y(x,y + D) = -H_y(x,y)$] and $\Re[{\cal H}(x+iD/2) = H_y(x,D/2) = 0$.

We seek the solution for which, when a current $I$ is first applied to the strips in the stack, the sheet-current density $K_z$ is equal to $K_c$ within bands of width $(a-c)$ at the edges but obeys $K_z < K_c$ in the middle region, $|x|<c.$  The simplest way to obtain this solution is to change variables as follows: 
$\tilde \zeta = (D/\pi)\sinh(\pi \zeta  /D)$, 
 $\tilde u = (D/\pi)\sinh(\pi u/D)$, 
$\tilde a = (D/\pi)\sinh(\pi a/D)$, and $\tilde c = (D/\pi)\sinh(\pi c/D)$, similar to the procedure used by Mawatari 
in Refs.\ \onlinecite{Mawatari96}, \onlinecite{Mawatari97b}, and \onlinecite{Mawatari99}.  Using $\tilde K_z(\tilde u) = K_z(u)$ and $\tilde {\cal H}(\tilde \zeta) = {\cal H}( \zeta)$, this yields
\begin{equation}
   \tilde {\cal H}(\tilde \zeta)= \frac{1}{2\pi} 
     \int_{-\tilde a}^{+\tilde a}d\tilde u \frac{\tilde K_z(\tilde u) \tilde u}{\tilde \zeta^2-\tilde u^2}. 
\label{singlestrip}
\end{equation}
This is the Biot-Savart law for an isolated strip carrying a current density $\tilde K_z(\tilde u) = \tilde K_z(-\tilde u),$ for which the solution is known to be\cite{Norris70,Brandt93,Zeldov94}

\begin{equation}
 \tilde {\cal H}(\tilde \zeta)= \frac{K_c}{\pi} \tanh^{-1}\sqrt{ \frac {\tilde a^2-\tilde c^2}{\tilde \zeta^2-\tilde c^2}},
\end{equation}
where here and in later similar expressions $\sqrt {\tilde \zeta^2-\tilde c^2}$ is shorthand for $(\tilde\zeta-\tilde c)^{1/2}(\tilde\zeta+\tilde c)^{1/2}$.  Thus the desired complex field is
\begin{equation}
{\cal H}( \zeta)= \frac{K_c}{\pi} \tanh^{-1}
\sqrt \frac {\sinh^2(\pi a/D)-\sinh^2(\pi c/D)}
{\sinh^2(\pi \zeta/D)-\sinh^2(\pi c/D)}.
\label{H}
\end{equation}
The corresponding complex potential,
\begin{equation}
{\cal G}( \zeta)= \int_{iD/2}^\zeta {\cal H}( \zeta')d\zeta',
\label{G}
\end{equation}
can be evaluated numerically.  Contours of constant $\Re {\cal G}( x+iy)$ (see Fig.\ 2) correspond to magnetic field lines.

\begin{figure}%***** Fig.2 ************************
\includegraphics[width=8cm]{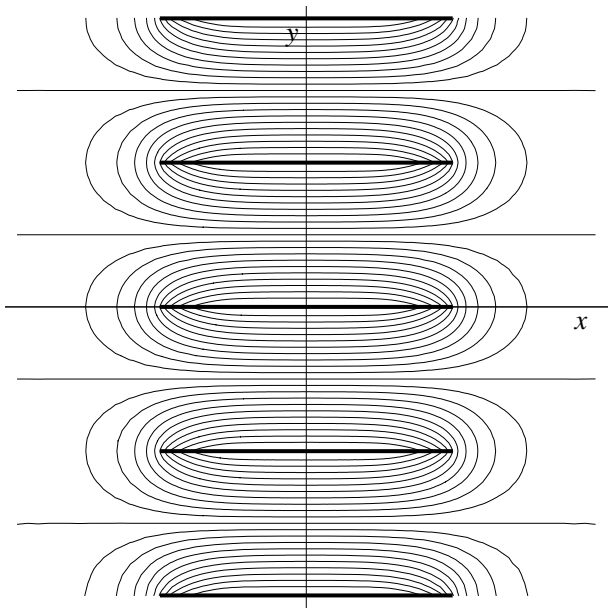}
\caption{Contours of constant $\Re  {\cal G}( x+iy)$ calculated from Eqs.\ (\ref{H}) and (\ref{G}) for a stack of thin superconducting strips of width $2a$ (thick lines) and spacing $D$.  The contours correspond to magnetic field lines, and in this figure when the current in is the $+z$ direction, the field lines circulate in a counterclockwise direction around the strips centered at $(x,y) = (0,-2D)$, $(0,0)$, and $(0,2D)$, and in a clockwise direction around the strips centered at $(0,-D)$ and $(0,D)$.  Here, $D = a = 2c$.   }
\label{Fig02}
\end{figure}

Taking the real and imaginary parts of ${\cal H}(x\pm i \epsilon)$ in Eq.\ (\ref{H}) and using $K_z(x) = H_x(x-i\epsilon)-H_x(x+i\epsilon)=2H_x(x-i\epsilon)$ yields
\begin{eqnarray}
H_y(x,0) & = & 0, \;|x| \le c, \label{Hy1} \\
 & = & \frac{K_c}{\pi} \tanh^{-1}
\sqrt \frac{\sinh^2(\pi x/D)-\sinh^2(\pi c/D)}{\sinh^2(\pi a/D)-\sinh^2(\pi c/D)}, \nonumber \label{Hy2} \\
&&\;\;\;\;\;\; c < |x| < a,\\
 & = & \frac{K_c}{\pi} \tanh^{-1}
\sqrt \frac{\sinh^2(\pi a/D)-\sinh^2(\pi c/D)}{\sinh^2(\pi x/D)-\sinh^2(\pi c/D)},\nonumber \label{Hy3} \\
&& \;\;\;\;\; \; |x| > a,\\
K_z(x)&=&\frac{2K_c}{\pi} \tan^{-1}
\sqrt \frac{\sinh^2(\pi a/D)-\sinh^2(\pi c/D)}{\sinh^2(\pi c/D)-\sinh^2(\pi x/D)},\nonumber \label{Kz1} \\
&& \;\;\;\;\; \; |x| \le c,\\
&=&K_c, \; c\le |x| < a. 
\label{Kz2} 
\end{eqnarray}

\begin{figure}%***** Fig.3 ************************
\includegraphics[width=8cm]{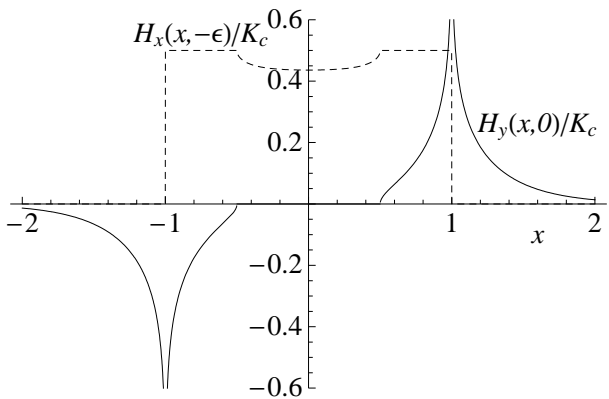}
\caption{Real (solid) and imaginary (dashed) parts of $\Re  {\cal H}( x-i\epsilon)= H_y(x,0)
+iH_x(x,-\epsilon)$ calculated from Eq.\ (\ref{H}) or Eqs.\ (\ref{Hy1})-(\ref{Kz2}) for a stack of thin superconducting strips of width $2a$ and spacing $D$.   Here, $a = 1$, $c = 0.5$, and $D = 1$. }
\label{Fig03}
\end{figure} 

The relationship between $c/a$ and the current $I$ carried by one of the strips is determined by the integral $I= \int_{-a}^{a}K_z(x) dx$, which can be expressed as
\begin{equation}
\frac{I}{I_c}=1-\frac{2}{\pi a}\int_0^c 
\tan^{-1}
\sqrt \frac{\sinh^2(\pi c/D)-\sinh^2(\pi x/D)}{\sinh^2(\pi a/D)-\sinh^2(\pi c/D)}dx,
\label{cvsI}
\end{equation}
where $I_c = 2aK_c$.
Shown in Fig.\ 4 are plots of $c/a$  vs  $(I/I_c)^2$ for $D/a =$ 0.1, 0.3, 1, 3, and 10.  For $D/a = 10$, the plot of $c/a$  vs $(I/I_c)^2$ is nearly indistinguishable from $c/a = \sqrt{1-(I/I_c)^2}$, the result known for an isolated strip.\cite{Norris70,Brandt93,Zeldov94}   On the other hand, for small values of $D/a$, the plot of $c/a$  vs $(I/I_c)^2$ can be calculated to good approximation by
\begin{equation}
\sinh(\pi c/D)=\sinh(\pi a/D)\cos[(\pi/2)(I/I_c)],
\label{Fapprox}
\end{equation}
such that the value of $c/a$ is very close to 1 except when $I/I_c$ is very close to 1. This behavior occurs because the current density $K_z(x)$ in the region $|x| < c$ is practically constant and nearly equal to $I/2a$  for a wide range of subcritical values of $I$, as shown by the plot of $K_z(x)$ vs $x$ for $D/a = 0.1$ in Fig. 5.
The curve of  $c/a$ vs $(I/I_c)^2$  for $D/a =$ 0.1 is indistinguishable from that obtained from Eq.\ (\ref{Fapprox}).  
For all values of $D/a$, $\delta =(a-c)=\alpha a (I/I_c)^2$ for small values of  $(I/I_c)$, where Eq.\ (\ref{cvsI}) yields 
\begin{equation}
\alpha = \frac{\pi a/D}{2\tanh(\pi a/D)}\Big[\frac{\pi/2}{{\bm K}(k)}\Big]^2,
\label{amc}
\end{equation}
and ${\bm K}(k)$ is the complete elliptic integral of the first kind of modulus $k=\tanh(\pi a/D)$.   This behavior is illustrated by the dashed lines in Fig.\ 4.  In the limit as $D/a \to \infty$, Eq.\ (\ref{amc}) yields $\alpha = 1/2$, and for $D/a \ll 1$, $\alpha = \pi D/8a$.

\begin{figure}%***** Fig.4 ************************
\includegraphics[width=8cm]{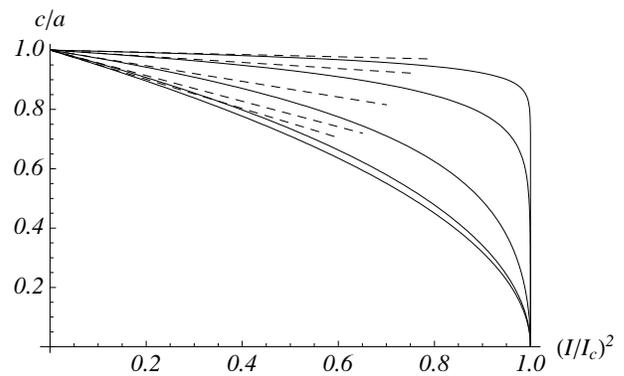}
\caption{Plots of   $c/a$ vs $(I/I_c)^2$ (solid) determined from Eq.\ (\ref{cvsI}) for $D/a$ = 0.1, 0.3, 1, 3, and 10 (top to bottom).  The dashed lines show the corresponding linear slopes for $(I/I_c)^2 \ll 1$ [Eq. (\ref{amc})]. }
\label{Fig04}
\end{figure} 

\begin{figure}%***** Fig.5 ************************
\includegraphics[width=8cm]{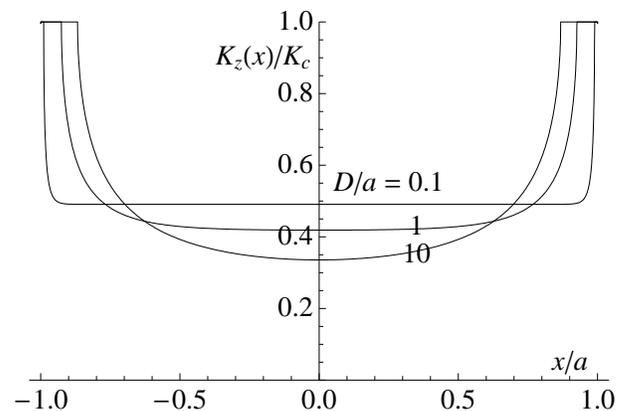}
\caption{Plots of $K_z(x)/K_c$ vs $x/a$ for $I/I_c = 0.5$ determined from Eqs.\ (\ref{Kz1}), (\ref{Kz2}), and (\ref{cvsI}) for $D/a$ = 0.1 ($c/a = 0.989$), 1 ($c/a = 0.926$), and 10 ($c/a = 0.868$).}
\label{Fig05}
\end{figure} 

\section{ac losses}

A secondary goal in this paper is to calculate the hysteretic ac losses in a bifilar stack of superconducting strips at a frequency $f = 1/T$ that is sufficiently low that eddy-current losses are negligible and the losses can be calculated using a quasistatic approach.\cite{Norris70}  The solutions for ${\bm H}(x,y)$ derived in Sec.\ II 
can be used to calculate $Q'$, the energy dissipated per cycle per unit length in each strip.  Consider time $t=0$, when the current $I$ has its maximum amplitude in the $z$ direction, the magnetic-field distribution is given by ${\bm H}(x,y) = \hat x H_x(x,y) + \hat y H_y(x,y)$, and the magnetic induction is ${\bm B}(x,y,0) = \mu_0 {\bm H}(x,y)$.  Half a cycle earlier, at time $t = -T/2$, when the current is in the opposite direction, ${\bm B}(x,y,-T/2) = -\mu_0 {\bm H}(x,y)$. The loss per cycle per unit length $Q'$ is twice the loss in the half cycle $-T/2 \le t \le 0.$ Thus 
\begin{equation}
Q' = 2\int_{-T/2}^0 \!\!dt \int_{-a}^a \!\!dx \int_{-d/2}^{d/2}\!\!dyJ_z(x,y,t)E_z(x,y,t). 
\label{Q'1}
\end{equation} 
According to critical-state theory,\cite{Campbell72} during this time interval, $E_z$ is nonzero only where $J_z$ is just above $J_c$, such that $J_z$ can be replaced by $J_c$ in Eq.\ (\ref{Q'1}), but the integral is to be carried out only over those portions of the cross section where $E_z(x,y,t) > 0$.  Note that $E_z(0,0,t) = 0$ when the current amplitude $I$ is less than $I_c$.  Next, let us use Faraday's law in the form $\oint d{\bm l} \cdot {\bm E} = - \int  d{\bm S} \cdot \partial {\bm B}/\partial t$, where the surface ${\bm S}$ consists of two rectangular parts of length $L_z$, one with width $x$ extending from the origin to $(x,0)$ and the other of width $y$ extending from $(x,0)$ to $(x,y)$.  Integration of Faraday's law thus yields
\begin{equation}
E_z(x,y,t)\! = \!\int_0^x \!\!dx' \frac{\partial B_y(x',0,t)}{\partial t}
-\int_0^y \!\!dy' \frac{\partial B_x(x,y',t)}{\partial t}.
\label{Ez}
\end{equation}
Substituting this expression into Eq.\ (\ref{Q'1}), integrating over time, noting that ${\bm B}(x,y,0)-{\bm B}(x,y,-T/2) = 2\mu_0 {\bm H}(x,y)$, and making use of the symmetry that the losses in the left and right halves of the strip are the same, we obtain 
\begin{eqnarray}
Q' &=& 8\mu_0 J_c \int_0^a dx \int_{-d/2}^{d/2} dy \int_0^x dx' H_y(x',0) \nonumber \\
&&-8\mu_0 J_c \int_0^a dx \int_{-d/2}^{d/2} dy \int_0^y dy' H_x(x,y')
\nonumber \\
&=&Q_e' +Q_{tb}'.
\label{Q'2}
\end{eqnarray}

From this expression we see that there are in general two important contributions to $Q'$ in superconducting strips.  The first of these is $Q_e'$, the dissipation due to magnetic flux in the form of vortex or antivortex segments transporting flux density $\mu_0 H_y$ in from the edges of the strip. The second  contribution is $Q_{tb}'$, the dissipation due to magnetic flux in the form of vortex or antivortex segments  transporting flux density $\mu_0 H_x$ in from the top and bottom surfaces of the strip. 

\subsection{Edge losses}

\begin{figure}%***** Fig.6 ************************
\includegraphics[width=8cm]{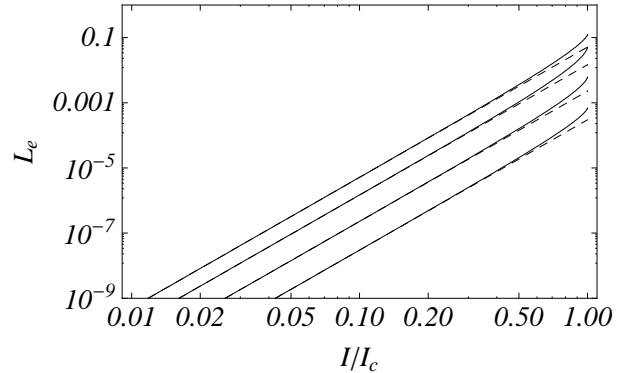}
\caption{Plots of the dimensionless function describing edge losses $L_e$ vs $F=I/I_c$ for $D/a$ = 0.1, 0.3, 1, and 10 (bottom to top).  The solid curves were obtained from Eq.\ (\ref{Le}) and the dashed lines from Eqs.\ (\ref{amc}) and (\ref{Leapprox}).  On this scale the solid curve for $D/a = 10$ is indistinguishable from  the Norris result $L_2$ [Eq.\ (\ref{L2})] for an isolated strip.\cite{Norris70} }
\label{Fig06}
\end{figure} 

The edge losses $Q_{e}'$ are most conveniently calculated from the fundamental equation\cite{Mawatari97a,Muller97,Mawatari99,Mawatari06,Claassen06,Mawatari07,
Norris70,Halse70}  
\begin{equation}
Q_{e}' = 8 \mu_0 K_c \int_c^a (a-x) H_y(x,0),
\label{Qebasic}
\end{equation}
which is obtained from the first term in Eq.\ (\ref{Q'2}) by partial integration.
Using the expression for $H_y(x,0)$ from Eq.\ (\ref{Hy2}),  we obtain 
\begin{equation}
Q_{e}' = \mu_0 I_c^2 L_e(F),
\label{Qe'}
\end{equation}
where $F = I/I_c$ and the dimensionless function $L_e$ is 
\begin{equation}
L_e=\frac{2}{\pi a^2}\int_c^a (a-x)\tanh^{-1}\!\!
\sqrt \frac{\sinh^2\frac{\pi x}{D}-\sinh^2\frac{\pi c}{D}}
{\sinh^2\frac{\pi a}{D}-\sinh^2\frac{\pi c}{D}}dx,
\label{Le}
\end{equation}
which is plotted as the solid curves in Fig.\ 6 for $D/a$ = 0.1, 0.3, 1, and 10.

In the limit as $D/a \to \infty$, the results correspond to those for an isolated strip, and $L_e(F)$ reduces to the function $L_2(F)$ derived by Norris,\cite{Norris70}
\begin{equation}
L_2(F) \!= \!\frac{(1\!-\!F)\ln(1\!-\!F)\!+\!(1\!+\!F)\ln(1\!+\!F)\!-\!F^2}{\pi}.
\label{L2}
\end{equation}

The maximum edge losses occur when $F = 1$ or $I = I_c$.  These can be calculated by setting $c = 0$ in Eq.\ (\ref{Le}).  In the limit as $D/a \to \infty$, this equation yields 
\begin{equation}
L_e(1)=(2\ln 2-1)/\pi = 0.123,
\label{Le1}
\end{equation}
as obtained by Norris for an isolated strip.\cite{Norris70}
For $D/a \ll 1$, on the other hand, Eq.\ (\ref{Le}) yields 
\begin{equation}
L_e(1)=\frac{7\zeta(3)}{4\pi^3}\Big(\frac{D}{a}\Big)^2= 0.0678\Big(\frac{D}{a}\Big)^2.
\label{Le1smallD}
\end{equation}
The smallness of this result arises because $H_y(x,0) \approx (K_c/\pi)\tanh^{-1}\{\exp[-\pi(a-x)/D]\}$ when $\pi a/D \gg 1$, such that, except for the logarithmic divergence at $x = a$, $H_y(x,0)$ is exponentially small, i.e., $H_y(x,0) \approx (K_c/\pi)\exp[-\pi(a-x)/D]$, over most of the range of integration in Eq.\ (\ref{Le}). 
The following interpolation function between large and small values of $D/a$ approximates $L_e(1)$ with a maximum error of 5\%:
\begin{equation}
L_e(1) \approx \frac{0.123}{[1+2.1(a/D)^{5/2}]^{4/5}}.
\label{Le1interp}
\end{equation}

Expanding the right-hand side of Eq.\ (\ref{Le}) to lowest order in powers of $F = I/I_c$ yields the approximation 
\begin{equation}
L_e(F) \approx \frac{2}{3\pi} \alpha^2 F^4,
\label{Leapprox}
\end{equation} 
where $\alpha$ is given by Eq.\ (\ref{amc}).  For large values of $D/a$, this equation  yields 
\begin{equation}
L_e(F) \approx F^4/6\pi,
\label{LelargeD}
\end{equation}
as found by Norris,\cite{Norris70}  and for small values of $D/a$, Eq.\ (\ref{Leapprox}) yields 
\begin{equation} 
L_e(F) \approx \frac{\pi}{96}\Big(\frac{D}{a}\Big)^2 F^4.
\label{LesmallD}
\end{equation}
Note from the solid curves and dashed lines in Fig.\ 6 that the approximation given in Eqs.\ (\ref{amc}) and (\ref{Leapprox}) provides a good estimate of the edge losses over a remarkably large range of values of $F = I/I_c$.

\subsection{Top-and-bottom losses}

In the theoretical analysis of the ac losses in thin films,\cite{Norris70,Brandt93} usually only  the losses due to vortex and antivortex motion in from the edges are taken into account.  These losses, represented by the term $Q'_e$, dominate in  isolated films when the ac current amplitude $I$ approaches $I_c$, for then the entering vortices travel an appreciable fraction of the strip width $2a$ during each cycle. See Eq.\ (\ref{Qebasic}).  On the other hand, vortices and antivortices entering from the top and bottom of the strips can travel at most a distance 
$d/2$, and when $d \ll a$, it makes sense to ignore the top-and-bottom losses, represented by the term $Q'_{tb}$. However, as seen in the above section, the edge losses are proportional to $F^4$ and also are much reduced for small values of $D/a$.  Since for small $F$  the top-and-bottom losses vary as a lower power of $F$, it is important to determine the conditions under which these losses exceed the edge losses.

To evaluate $Q_{tb}'$, we make use of the fact that, although in general $J_z(x,y) = \partial H_y(x,y)/\partial x-\partial H_x(x,y)/\partial y,$ in thin films the second term is far larger in magnitude. Therefore, to excellent approximation,
\begin{eqnarray}  
H_x(x,y)&=&-J_c(y-y_p),\;y_p < y <d/2,
\\
&=&0 ,\; -y_p < y < y_p, \\
&=&-J_c(y+y_p),-d/2< y < -y_p ,
\end{eqnarray}
where $H_x(x,-y) = -H_x(x,y)$.  Here $y_p(x) = d/2-H_x(x,-d/2)/J_c$ for $|x|<c$, where $H_x(x,-d/2)<J_cd/2$, or $y_p(x) = 0$ for $c \le |x| < a$, where $H_x(x,-d/2)=J_cd/2$.    Carrying out the second integral in Eq.\ (\ref{Q'2}), we obtain
\begin{equation}
Q_{tb}' = \frac{8\mu_0}{3J_c}\int_0^a |H_x(x,\pm d/2)|^3 dx.
\label{Qtbintegral}
\end{equation}
This result is expected, because when a type-II superconductor is subjected to a parallel ac field of amplitude $H_0$, the hysteretic ac loss per unit area per cycle is known to be\cite{Campbell72}
\begin{equation}
Q_A = \frac{2\mu_0 H_0^3}{3J_c}.
\label{QA}
\end{equation}
The result for $Q_{tb}'$ in Eq.\ (\ref{Qtbintegral}) corresponds to replacing $H_0$ by $|H_x(x,\pm d/2)|$ and integrating Eq.\ (\ref{QA}) over the top and bottom of the film.  Since we are here considering the case that $d \ll a$, we can evaluate Eq.\ (\ref{Qtbintegral}) by replacing  $|H_x(x,\pm d/2)|$ by $H_x(x,-\epsilon) = K_z(x)/2$, where $K_z(x)$ is given by Eqs.\ (\ref{Kz1}) and (\ref{Kz2}).
The top and bottom losses therefore can be expressed as 
\begin{equation}
Q_{tb}'=\mu_0 I_c^2 L_{tb}(F),
\label{Qtb}
\end{equation}
where $F = I/I_c$, and the dimensionless function $L_{tb}$ is
\begin{eqnarray}
L_{tb}&=&\frac{d}{12 a}\Big\{1-c/a \nonumber \\
&+&\!\!\!\frac{8}{\pi^3 a}\!\int_0^c \!\Big[\tan^{-1}\!\sqrt{\frac{\sinh^2\frac{\pi a}{D}\!-\!\sinh^2\frac{\pi c}{D}}{\sinh^2\frac{\pi c}{D}\!-\!\sinh^2\frac{\pi x}{D}}}\Big]^3dx\Big\}.
\label{Ltb}
\end{eqnarray}
The solid curves in Fig.\ 7 show plots of $L_{tb}$ vs $I/I_c$ for $D/a$ = 0.1, 0.3, 1, and 10 for the example of $d/a = 0.001$. 
All the curves meet at $F=1$ or $I=I_c$, where $c = 0$, such that
\begin{equation}
L_{tb}(1) = \frac{d}{12a}.
\label{Ltb1}
\end{equation}

\begin{figure}%***** Fig.7 ************************
\includegraphics[width=8cm]{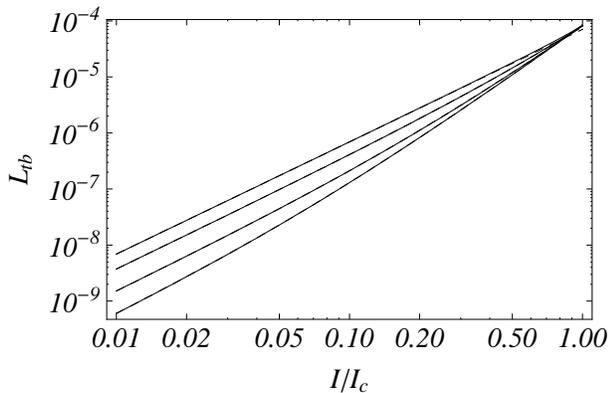}
\caption{Plots of the dimensionless function describing top-and-bottom losses $L_{tb}$ vs $F=I/I_c$ for $d/a = 0.001$ and $D/a$ = 0.1, 0.3, 1, and 10 (bottom to top).  The solid curves were obtained from Eq.\ (\ref{Ltb}), and the corresponding dashed curves, which on this scale are indistinguishable from the solid curves except for $D/a = 10$, were obtained from Eqs.\ (\ref{amc}), (\ref{Ltbapprox}), and (\ref{beta}).  }
\label{Fig07}
\end{figure} 

Expanding the right-hand side of Eq.\ (\ref{Ltb}) through third order in powers of $F = I/I_c$  leads to the approximation
\begin{equation}
L_{tb} \approx \frac{d}{12 a}(1.6855 \alpha F^2 + \beta F^3),
\label{Ltbapprox}
\end{equation} 
where part of the first term, $1.0000\alpha F^2$, comes from the term $1-c/a$, and the other part, $0.6855 \alpha F^2$, comes from expansion of the integral in Eq.\ (\ref{Ltb}) and the result
\begin{equation}
\frac{16}{\pi^3}\int_0^\infty(\tan^{-1}u)^3/u^3 du = 0.6855.
\end{equation}
The factor $\beta$ in Eq.\ (\ref{Ltbapprox}) is given by 
\begin{equation}
\beta=\frac{8}{\pi^2}\frac{(\pi a/D)}{\tanh(\pi a/D)}
\Big[1-\frac{{\bm E}(k)}{{\bm K}(k)}\big]\alpha,
\label{beta}
\end{equation}
where ${\bm K}(k)$ and ${\bm E}(k)$ are complete elliptic integrals of the first and second kind of modulus $k = \tanh(\pi a/D)$.  
When $D/a \gg 1$, $\beta\approx 2(a/D)^2$, and in the limit $D/a \to 0$, $\beta \to 1$.  
The dashed curves in Fig.\ 7 show plots of $L_{tb}$ obtained from Eq.\ (\ref{Ltbapprox}) for $D/a$ = 0.1, 0.3, 1, and 10 (bottom to top). 
The figure shows that this approximation is excellent for all $F = I/I_c$ except for $D/a = 10$  close to $F=1$.
In the limit $D/a \to 0$, when $\alpha \to 0$ and $\beta \to 1$, the approximation in Eq.\ (\ref{Ltbapprox}) becomes exact, yielding
\begin{equation}
L_{tb} = \frac{d}{12a}F^3,
\label{LtbsmallD}
\end{equation}
which holds for all $F$.

\subsection{Comparison of edge and top-and-bottom losses}

The total energy dissipated per cycle per strip per unit length is $Q' = Q_e' + Q_{tb}' =\mu_0 I_c^2 L(F),$ the sum of the edge and top-and-bottom losses. Figure 8 shows plots of $L=L_e +L_{tb}$ vs $F = I/I_c$ for $D/a$ = 0.1, 0.3, 1, and 10 and $d/a = 0.001$, the cases considered in Figs.\ 6 and 7.  It is useful to define $F_X$ as the value of $F$ where the edge and top-and-bottom losses are equal, i.e., where the curves of $L_e$ vs $F$ and $L_{tb}$ vs $F$ cross. 
Equations (\ref{Le}) and (\ref{Ltb}) yield for $d/a = 0.001$ the values $F_X =$ 0.304, 0.096, 0.051, and 0.037 for $D/a =$ 0.1, 0.3, 1, and 10.
The solid portions of the curves in Fig.\ 8 show where $F > F_X$ and $L_e > L_{tb}$, and the dashed portions show where $F < F_X$ and $L_e < L_{tb}$.  The edge losses are most important when $D/a$ is large, and the top-and-bottom losses grow in relative importance when $D/a$ is small.
The dotted curves show $L = L_e+L_{tb}$ calculated from the power-law approximations given in Eqs.\  (\ref{Leapprox}) and (\ref{Ltbapprox}).

\begin{figure}%***** Fig.8 ************************
\includegraphics[width=8cm]{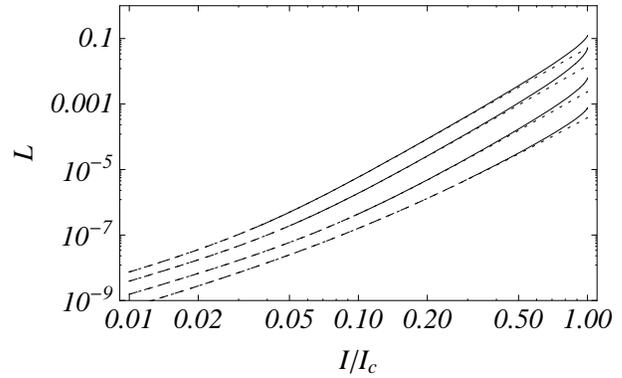}
\caption{Plots of the dimensionless function $L=L_e+L_{tb}$ describing the sum of the edge losses $L_e$ [Eq.\ (\ref{Le})] and top-and-bottom losses $L_{tb}$ [Eq.\ (\ref{Ltb})] vs $F=I/I_c$ for $D/a$ = 0.1, 0.3, 1, and 10 (bottom to top) and $d/a$ = 0.001.  The solid curves display those portions of $L$ for which $F > F_X$ and $L_e > L_{tb}$, and the dashed curves display those portions for which $F < F_X$ and $L_e < L_{tb}$. The dotted curves show $L$ calculated from the sum of the approximate expressions for $L_e$ [Eq.\ (\ref{Leapprox})] and $L_{tb}$ [Eq.\ (\ref{Ltbapprox})]. }
\label{Fig08}
\end{figure} 

\begin{figure}%***** Fig.9 ************************
\includegraphics[width=8cm]{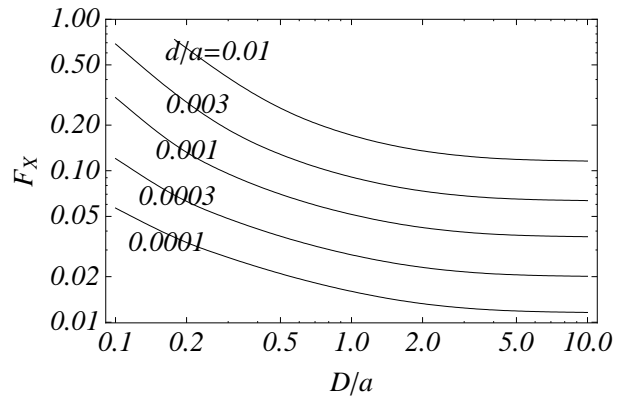}
\caption{Plots of $F_X$, the value of $F = I/I_c$ for which the edge losses [Eq.\ (\ref{Le})] are equal to the top-and-bottom losses [Eq.\ (\ref{Ltb})], vs $D/a$ for several values of $d/a$.  For given values of $D/a$ and $d/a$, when $F>F_X$,  edge losses exceed  top-and-bottom losses, but when $F<F_X$,  top-and-bottom losses exceed  edge losses.}
\label{Fig09}
\end{figure} 

Figure 9 displays plots of $F_X$ vs $D/a$ for values of $d/a$ = 0.0001, 0.0003, 0.001, 0.003, and 0.01.
These plots show that when $D/a$ is small enough and $d/a$ is large enough, the top-and-bottom losses exceed the edge losses for all values of $F = I/I_c$.

\section{Discussion}

In Sec.\ II of this paper, I derived general expressions for the magnetic-field and sheet-current-density distributions for an infinite bifilar stack of thin superconducting strips of width $2a$, thickness $d$, and separation $D$, all carrying current of magnitude $I$, but with adjacent strips carrying current in opposite directions.  The calculations assumed that ${\bm B} = \mu_0 {\bm H}$ and that the critical current density $J_c$ of each strip was uniform and independent of the local magnetic flux density.

In Sec.\ III, I used critical-state theory to derive expressions for $Q'$, the hysteretic ac loss per cycle per unit length  in each tape, where $Q'$ is the sum of edge losses $Q'_e$ and top-and-bottom losses $Q'_{tb}$.  The  top-and-bottom losses grow in relative importance as $D$ becomes less than $a$.  I expressed $Q'$ in terms of integrals of the $x$ and $y$ components of the magnetic  fields given in Sec.\ II.  Although these integrals can easily be evaluated by numerical integration, I also expanded $Q'$ in powers of  $F = I/I_c$ to obtain some useful analytic approximations.

The behavior of the losses in a bifilar stack are very different from those in an infinite stack of strips of separation $D$ when the strips all carry the same current in the {\it same} direction.  In the latter case,
$Q'$, the hysteretic loss per unit length per cycle in each tape,  increases rapidly as $D/a$ decreases.\cite{Mawatari97a,Muller97}  The additive effect of the magnetic fields generated by nearby strips greatly increases the magnetic field at the edges of a given strip, such that the edge losses are greatly magnified.  When $D/a \ll 1$ and $F = 1$ ($I = I_c$), the resulting $Q'(1)$ is then larger than that of an isolated strip by a factor of $2.71(a/D)$.
For the case of a finite stack, a similar enhancement of the losses over those in an isolated strip has been noted experimentally\cite{Grilli07} and theoretically.\cite{Grilli07,Clem07}

In strong contrast, in a bifilar stack, where adjacent strips carry current in {\it opposite} directions, the magnetic fields generated by nearby strips nearly cancel, and when $D/a \ll 1$, the magnetic-field distribution is strongly altered by the presence of adjacent strips.  The perpendicular component of the field is strongly attenuated, and the parallel component becomes nearly uniform across the width of the strip.  
In this case, accounting for both the edge and top-and-bottom losses in the stack, $Q'(1)$ is smaller than that for an isolated strip by a factor of $0.552(D/a)^2+0.678(d/a)$ [see Eqs.\ (\ref{Le1}),  (\ref{Le1smallD}), and (\ref{Ltb1})].  

Majoros et al.\cite{Majoros07} used a finite-element method to calculate the transport ac losses in finite stacks of superconducting tapes of elliptical cross section carrying mutually antiparallel currents at the critical value $I_c$.  They found that when the tapes were closely spaced, the losses were less than when the tapes were far apart.  The results in Sec.\ III confirm this general conclusion not only for $I= I_c$ but also for all $I < I_c$.

Although  the results  in Sec.\ III for $Q'$ were derived for an infinitely tall bifilar stack of infinitely long strips, they should provide an excellent approximation to the hysteretic loss per cycle per unit length for superconducting tapes of finite length in fault-current limiters consisting of noninductively wound pancake coils with a large number of bifilar windings, so long as the radius $r$ of each winding is much larger than the spacing $D$ between adjacent tapes.

A key assumption made in this paper  is that the magnetic induction in the superconducting film is given by ${\bm B} = \mu_0 {\bm H}$.  This should be a good approximation in a high-$\kappa$ superconducting film when the magnitude of the self-field $\bm H$ at the edges or surfaces is much larger than the lower critical field $H_{c1}$.  However, if this condition is not met, then the loss expressions given in Sec.\ III will generally overestimate  the true losses.  For example, for the case of $D/a \ll 1$, when $K_z \approx I/2a$, if $H_x(x,-\epsilon) = K_z/2 \approx I/4a$ is less than $H_{c1}$, no vortices will be able to penetrate the top or bottom of the strip.  In this case, the top-and-bottom losses will be zero, rather than what was calculated in Sec.\ III B.

\begin{acknowledgments} I thank A. P. Malozemoff for posing questions that
stimulated my work on this problem.  Work at the Ames Laboratory was
supported by the Department of Energy - Basic Energy Sciences under Contract
No. DE-AC02-07CH11358.
\end{acknowledgments}

\end{document}